\documentclass[letterpaper, 10 pt, conference]{ieeeconf} 
\usepackage[style=ieee,backend=biber, url=false, doi=false, isbn=false, maxnames=100]{biblatex}

\usepackage{graphicx} %use graph format
\usepackage{epstopdf}
\usepackage{flushend}

\usepackage{xcolor}
\usepackage{amsmath,array,graphicx}
\usepackage{kantlipsum}
\usepackage{latexsym}
\usepackage{amsfonts}
\usepackage{amssymb}
\usepackage{multirow}
\usepackage{gensymb}
\usepackage{indentfirst}

\newcommand{\startpara}[1]{{\vskip1pt\noindent{\bf #1.}}}
\title{\LARGE \bf  A Study on Learning and Simulating \\ Personalized Car-Following Driving Style

}

\author{Shili Sheng$^{1}$, Erfan Pakdamanian$^{1}$, Kyungtae Han$^{2}$, Ziran Wang$^{2}$, and Lu Feng$^{1}$
\thanks{$^{1}$Shili Sheng, Erfan Pakdamanian, and Lu Feng are with School of Engineering and Applied Science, University of Virginia, Charlottesville, VA 22904, USA
        {\tt\small \{ss7dr, ep2ca, lf9u\}@virginia.edu}}%
\thanks{$^{2}$Kyungtae Han and Ziran Wang are with Toyota Motor North America-InfoTech Labs, Mountain View, CA 94043, USA
        {\tt\small \{kyungtae.han, ziran.wang\}@toyota.com}}
}

\begin{document}
\maketitle
\thispagestyle{empty}
\pagestyle{empty}
%%%%%%%%%%%%%%%%%%%%%%%%%%%%%%%%%%%%%%%%%%%%%%%%%%%%%%%%%%%%%%%%%%%%%%%%%%%%%%%%

\begin{abstract}
Automated vehicles are gradually entering people's daily life to provide a comfortable driving experience for the users. The generic and user-agnostic automated vehicles have limited ability to accommodate the different driving styles of different users. This limitation not only impacts users' satisfaction but also causes safety concerns. Learning from user demonstrations can provide direct insights regarding users' driving preferences. However, it is difficult to understand a driver's  preference with limited data. In this study, we use a model-free inverse reinforcement learning method to study drivers' characteristics in the car-following scenario from a naturalistic driving dataset, and show this method is capable of representing users' preferences with reward functions. In order to predict the driving styles for  drivers with limited data, we apply Gaussian Mixture Models and compute the similarity of a specific driver to the clusters of drivers. We design a personalized adaptive cruise control (P-ACC) system through a partially observable Markov decision process (POMDP) model. The reward function with the model to mimic drivers' driving style is integrated, with a constraint on the relative distance to ensure driving safety. Prediction of the driving styles achieves 85.7\% accuracy with the data of less than 10 car-following events. The model-based experimental driving trajectories demonstrate that the P-ACC system can provide a personalized driving experience.

\end{abstract}
\section{Introduction}
Automated vehicles have attracted significant attention from the research community because of their promising ability to increase traffic safety and enhance the human driving experience for various driving situations. Most  automated driving technologies are currently at SAE level 2~\cite{bagloee2016autonomous} relying on the development of advanced driver-assistance systems~(ADAS)~\cite{shaout2011advanced}.  Car-following is a fundamental building block of automated longitudinal motion control, various ADAS are further developed to enable a safer driving experience, including  cruise control (ACC) and forward-collision warning \cite{wang2020asurvey}.

Most available ADAS are generic and user-agnostic, which limits their ability to fit drivers' different styles
while different drivers tend to have different driving styles~\cite{taubman2004multidimensional}. 
For example, an ACC can be too aggressive for a driver who prefers a large distance gap to preceding vehicles, while too conservative for a driver who prefers a smaller gap~\cite{Yi2019ARecognition}. Personalized ADAS aims to integrate drivers' preferences into the system design and adapt to diverse drivers' styles~\cite{Gao2020PersonalizedControl}. Because of the system ability to predict and mimic driving behavior, it can provide optimal driving experience, improve traffic safety~\cite{Zhu2018PersonalizedIdentification}, and enhance drivers' trust~\cite{Bolduc2019MultimodelControl}.

\begin{figure}
  	\centering
  	\includegraphics[width=70mm]{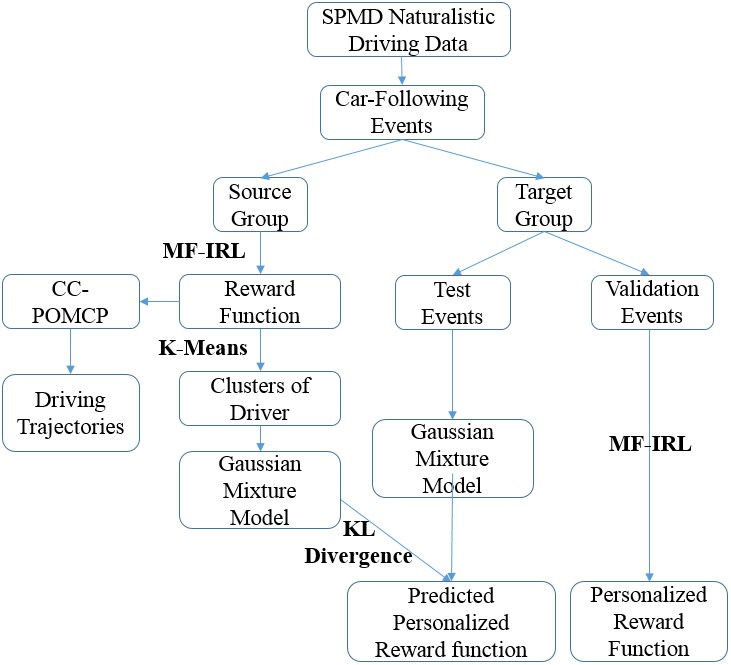}
  	\caption{Overview of the proposed method.}
  	\label{fig:overview}
\end{figure}

 Learning the individual driving styles allows to predict drivers' maneuvers, make a decision at the individual level, and facilitate the design of  personalized ADAS.  

Driving preferences can be learn through simply demonstrating driver's driving style ~\cite{silver2013learning},
rather than manually tuning various parameters to adapt to drivers' behaviors ~\cite{rosbach2019driving}
since this can be considered that drivers often maximize some reward in their mind and trade-off different factors, such as speed, acceleration, and distance to surrounding vehicles
in many Inverse Reinforcement Learning (IRL) setting ~\cite{abbeel2004apprenticeship}.  

%  sun2018probabilistic,sun2018courteous
 IRL was commonly used to infer the driving styles by learning the cost function that best explains the observed demonstrations \cite{sun2018probabilistic,liao2022online}. A cost function usually combines various features and weights~  \cite{Kuderer2015LearningDemonstration}. Using cost functions to represent driving styles can also facilitate predicting and imitating driver behaviors, and further generate comfortable motions and predictable behaviors~\cite{naumann2020analyzing}. Currently,  most IRL methods rely on the prior knowledge  of the transition model, and it is often difficult to satisfy in real life~\cite{arora2021survey}. Model-free IRL methods relax this requirement~\cite{uchibe2018model}, and it can achieve a better performance than traditional model-based IRL methods~\cite{vroman2014maximum}.

 In real-life situations, driving observations from one specific driver may be insufficient to develop individual-based personalized ADAS  \cite{lu2019transfer, li2019transferable, hasenjager2019survey}.
Group-based personalization studies drivers' behavior with a small number of representative styles \cite{hasenjager2019survey, wang2020driver}.  The challenge lies in how to adapt group-based personalization to individual-based personalization when only limited data from a specific driver is available.

In this paper, we present a personalized car-following style learning method for drivers with limited data as shown in Figure~ \ref{fig:overview}. We used a model-free IRL to learn each individual driver's style~(i.e., reward function) from 42 drivers and cluster them into four different groups. Furthermore, we developed Gaussian Mixture Models (GMM), used the Kullback–Leibler (KL)-Divergence to  measure the similarities, and predicted the personalized reward function. In order to validate this approach, we    developed a car-following model with a partially observable Markov decision process (POMDP) to mimic drivers' driving preferences with a constraint on the minimum distance between the vehicle  to ensure safety.

We highlight the following contributions in this work:
\begin{itemize}
\item The effectiveness of the model-free IRL  in learning driving preference from naturalistic driving data is demonstrated.
\item Driving preferences are clustered into four representative styles based K-means. 
\item Personalized driving styles are learned for drivers with limited data based on the KL-Divergence of GMMs.
\item  We designed a personalized ACC control utilizing the learned IRL reward function and POMDP with  a constraint on the minimum distance between the vehicle  to ensure safety.
\end{itemize}

The remainder of the paper is organized as follows: Section \ref{RelatedWork} summarizes the related work,   Section \ref{overview} shows an overview of our method,  Section  \ref{dataset} describes the dataset and the preprocessing, Section \ref{IRL} introduces the car-following preference learning based on the IRL, Section \ref{cluster} explains the driving styles clustering, Section \ref{prediction} presents the prediction using drivers' limited information, Section~\ref{POMCP} illustrates the design of the  POMCP model for the personalized ACC controller, and Section \ref{conclustion} draws the conclusion.
 \section{Related work}
 \label{RelatedWork}

\subsection{Personalized Advanced Driver-Assistance Systems}
Various personalized ADAS are proposed to assist the car-following scenarios. Group-based personalized ACC has been developed by assigning drivers into a small number of groups~\cite{hasenjager2019survey}. Gao et al. proposed a personalized ACC based on three driving styles learned from 66 drivers by using unsupervised clustering methods, and then the parameters representing each style are fed into a model predictive controller~(MPC) \cite{gao2020personalized}. Zhu et al. used KL-divergence to measure the similarity among 84 drivers and classified them into three groups ~\cite{Zhu2019Typical-driving-style-orientedData}. A personalized ACC was further designed to meet different groups' driving characteristics in speed control and distance control. Wang et al. developed a learning-based personalized driver model based on bounded generalized GMM for car-following scenarios \cite{Wang2018AModel}.  Chen et al.~\cite{chen2017learning} proposed a learning model for personalized ACC that can learn and replicate driver behaviors within acceptable errors. Wang et al. developed a Gaussian Process Regression algorithm for personalized ACC, where both numerical and human-in-the-loop experiments verify the effectiveness of the proposed algorithm in terms of reducing the interference frequency by the driver \cite{wang2021personalized} .

% \startpara{Reinforcement Learning}
\subsection{Inverse Reinforcement Learning}
Ng and Russell~\cite{ng2000algorithms} introduced an IRL to extract a reward function for observed optimal behaviors.  Abbeel  and  Ng~\cite{abbeel2004apprenticeship}  used an IRL to learn five different driving styles on a highway simulation, assuming the reward function can be expressed as a linear combination of known features. 
% \textcite{Kuderer2015LearningDemonstration}  
Kuderer et al.~\cite{Kuderer2015LearningDemonstration} introduced a feature-based IRL method for learning individual navigation styles from the real driving demonstrations on a highway. They demonstrated that their method was able to achieve distinct mean acceleration and jerk for two users. They concluded that distinct policies can be learned for different users. \cite{levine2012continuous}  learned different rewards for three different driving styles based on an IRL method using locally optimal demonstration.   Naumann et al.~\cite{naumann2020analyzing}  investigated the suitability of explaining human driving behavior with the cost functions. They explored various features such as longitudinal acceleration and longitudinal jerks from human driver trajectories. They inferred human drivers' preference over the features through the learned cost function weights. Jain et al. \cite{Jain2019Model-freeEstimation} proposed a model-free IRL using maximum likelihood estimation to investigate drivers' preferences  in the scenario of freeway merging based on 12 trajectories. The studies by Gao et al. \cite{gao2018car} and Zhao et al. \cite{zhao2022personalized} both show that their IRL algorithms can recover the personalized car-following gap preference based on different vehicle speed values, where a gap-speed matrix can be used to design the control logic for personalized ACC systems.
\section{Research Overview}
\label{overview}
In Figure~\ref{fig:overview}, we summarized our method for learning the personalized car-following preference for drivers with limited data. We extracted car-following events from a naturalistic driving dataset. We kept the 49 drivers with more than 10 car-following events. We considered 42 drivers with no more than 60 events as the \textit{source group}  and the seven drivers as the \textit{target group}. 

For the drivers in the source group, we employed a model-free IRL method to infer their driving styles through the reward functions.  We further clustered these driving styles into four representative styles using K-Means. 

In order to demonstrate how to infer the driving style of a driver with limited data, we separated 10 events as the \textit{test events} for each driver in the target group, and keep the rest as the \textit{validation events}. Assuming we would not be able to learn the  driving styles through the IRL directly with the limited data in the test events, we developed a GMM for them individually instead.  Then we computed the individual diver's similarity with the GMM models of the 4 clusters using KL-divergence, respectively. We inferred which cluster that the driver is most similar with according to the KL-divergence. Then we assigned that individual driver with the centroid reward function of that cluster. Thus, we were able to predict the personalized reward function with limited data. To validate the effectiveness of this method, we used the validation events to learn the reward function directly through the IRL as the ground truth. We demonstrated the prediction accuracy in Section~\ref{prediction}.

\section{Dataset}
\label{dataset}
\begin{figure}[t]
   	\centering
   	\includegraphics[width=55mm]{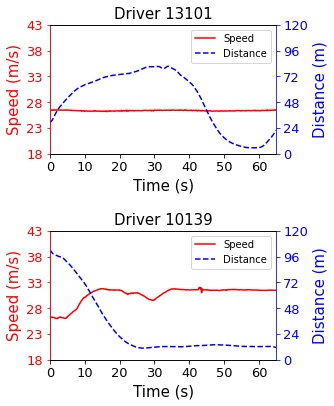}
   	\caption{Car-Following events.}
   	\label{fig:rawData}
\end{figure} 

We extracted car-following data from the datasets of \textit{DataWsu} and \textit{DataFrontTargets} in the project of Safety Pilot Model Deployment (SPMD)\footnote{https://catalog.data.gov/dataset/safety-pilot-model-deployment-data}. The SPMD project collected naturalistic driving data (i.e  without any restriction of particular routes or driving time) in Ann Arbor, Michigan, USA. 
Ninety-eight sedans integrated with data acquisition systems~(DAS) were provided to drivers. \textit{DataWsu} recorded data such as speed and acceleration via the vehicles' Controller Area Network (CAN) Bus at a frequency of 10 Hz. \textit{DataFrontTargets} recorded data such as relative distance and relative speed using the Mobileye at a frequency of 10 Hz. Each vehicle had a unique device ID (eg. 10205). We assumed that each vehicle was only assigned to one individual driver,  and we used the device ID as the driver ID. We obtained the car-following data for 85 drivers with the following criteria to select car-following events.  
\begin{itemize}
    \item The ego vehicle followed its closest preceding vehicle for at least 30 seconds. 
    \item The relative distance between the ego vehicle and preceding vehicle was always shorter than 120 meters. If the relative distance was longer than 120 meters, we assume that the vehicle was under free driving \cite{Wang2019AModels}.
    \item The speed of the ego vehicle was always between 18 $m/s$ and 43 $m/s$.
\end{itemize}

% \begin{figure}[t]
%   	\centering
%   	\includegraphics[width=\linewidth]{Figures/gmap.jpg}
%   	\caption{Driving scenario (Driver ID: 10106, Trip ID: 314).}
%   	\label{fig:gmap}
% \end{figure} 

In total, there were 36 drivers with no more than 10 car-following events, and 49 drivers with more than 10 events. We kept the 49 drivers for further study.

 Figure \ref{fig:rawData} shows two car-following  events: Driver 13101 was driving at relatively low speed ($26.4 \pm 0.08$~$m/s$) while the distance to the front vehicle was changing. The driver focused on keeping the speed stable while being indifferent to the changing distance. On the contrary, Driver 10139 increased the speed to shorten the distance, and then kept at a stable speed and distance. 
\section{Inverse Reinforcement Learning}
\label{IRL}
\subsection{Preliminaries}
% In this subsection, we introduce the model-free IRL method to study each driver's driving style.
% \subsubsection{Reinforcement learning}
% Reinforcement learning can usually be described by a Markov decision process using a tuple $<S, A, T ,R, \gamma>$. $S$ denotes the state space that the agent can perceive in the environment; $A$ denotes the action space that the agent can perform; $\mathcal{T}: S \times A \timeS \to \mathbb{R}$ denotes the transition probability; $R: S \to \mathbb{R}$ denotes the reward function; and $\gamma \in [0,1]$ denotes the discount factor, i.e. the greater the $\gamma$ is, the more the agent values the future rewards. At each time step $t$,  a state $s_t \in S$ transit to $s_{t+1} \in S$ with probability $\mathcal{T}(s_{t+1} | s_t, a_t)$ given an action $a_t \in A$, and the agent receives a reward $R(s_t, a_t)$. A RL policy $\pi$ maps a state $s$ to an action $a$. The RL task aims to compute the optimal policy $\pi^*$ to maximize the expectation of the cumulative reward $\mathbb{E} [\sum_{t=0}^\infty \gamma^t R(s_t, a_t)]$.
% \subsubsection{Inverse reinforcement learning}
% IRL aims to learn a policy that an agent adopts to act in an environment, whereas the reward function is not available, but successful demonstrations by an expert are given 

The reward function in reinforcement learning  determines the policy that the agent will adopt to act in an environment. However, the reward function is not always available. Instead, an expert's behavior is easier to observe. The IRL aims to derive the reward function from the observed behaviors~\cite{shiarlis2016inverse}. In the car-following scenario, different drivers have their own driving styles and value the  environmental factors differently. Drivers' driving styles determine their driving actions as the reward function determines the agent's policy. 
We can learn different drivers' driving styles using the IRL as we derive the reward function from the recorded car-following data.

The reward function is usually formed as a linear combination of binary features $\phi: S \times A \to \{ 0, 1\}$, where $S$ denotes the state space that the agent can perceive in the environment, and $A$ denotes the action space that the agent can perform. The reward function for expert $E$ can be denoted as $R_E(s,a)= \sum_{m=1}^{M} \omega_m \cdot \phi_m(s,a) $, where $M$ is the number of features and $\omega$ is the weight. Ultimately, the IRL is learning the weights such that the demonstrated behavior is optimal.

% $ \mu = \sum_{k=1}^{K} \phi_k(s,a) $ is also known as the feature expectation.

Model-based IRL methods such as Bayesian inference~\cite{ramachandran2007bayesian}, maximum entropy~\cite{ziebart2008maximum}, and maximum likelihood estimation~\cite{vroman2014maximum} require prior knowledge of transition function, which is not easy to obtain in real life. Jain et al. ~\cite{Jain2019Model-freeEstimation} proposed a model-free IRL method \textit{Q-averaging}  by estimating the Q-value without knowledge of the transition function and it achieves a higher log-likelihood compared with an existing model-based method. We employed the \textit{Q-averaging} IRL method in this study.
\begin{figure}[t]
   	\centering
   	\includegraphics[width=55mm]{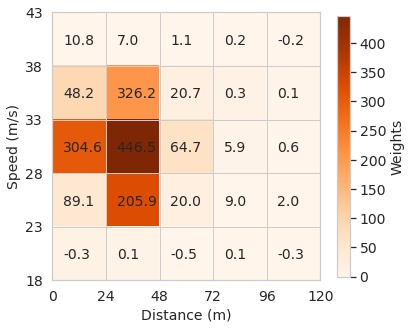}
   	\caption{Weights of the reward function for Driver 10575.}
   	\label{fig:reward_one}
\end{figure}

\subsection{Learning Car-Following Preference}
In order to capture the variability of the driving states, we aggregated data of every three seconds into one data point.  We summarized the setup of the car-following data with IRL format as follows. 
\startpara{State Space} We defined the state space with two state variables:
\begin{itemize}
    \item $d$: The relative distance from the ego vehicle to its closest preceding vehicle.
    \item $v$: The velocity of the ego vehicle from the vehicle’s CAN Bus.
\end{itemize}

We discretized the speed of the vehicles and their relative distance to front vehicles into five intervals evenly, which led to 25 states for each vehicle. 
\startpara{Action}
We modeled the acceleration (in $m/{s^2}$) as the actions of the drivers.
\begin{itemize}
    \item  High brake $(acc \leq -1.46 )$, \item Mild brake $(-1.46 < acc \leq -0.18 )$, \item Minimal acceleration $(-0.18 < acc \leq 0.18 )$, \item Mild acceleration $(0.18 < acc \leq 1.46 )$, \item High acceleration $(acc>1.46)$.
\end{itemize}
\startpara{Feature}
We used 25 features to indicate the states of relative distance and velocity of the vehicle.
\startpara{Reward Function}
Figure \ref{fig:reward_one} shows the learned feature weights for Driver 10575 using the model-free IRL approach. It explains the driver's preference over different driving states. The weight of the $12^{th}$ feature (speed between 28 $m/s$ and 33 $m/s$, distance between 24 $m$ and 48  $m$) is the greatest, corresponding to the state that Driver 10575 is most comfortable with. The cumulative weights over the states with the distance between 24 $m$  and 48 $m$ are greater than the weights over the states with speed between 28 $m/s$ and 33 $m/s$, indicating that he/she prefers maintaining a stable distance rather than maintaining a stable speed. The states with fewer weights (e.g., states with speed below 23 $m/s$, and states with distance above 96 $m$) show the driving situations that the driver prefers not to be in.

\section{Clustering}
\label{cluster}

\begin{figure}[t]
   	\centering
   	\includegraphics[width=55mm]{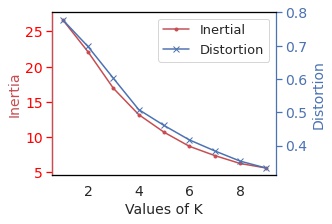}
   	\caption{Inertia and distortion with respect to the value of $K$.}
   	\label{fig:elbow}
\end{figure}

\begin{figure}[t]
  	\centering
  	\includegraphics[width=70mm]{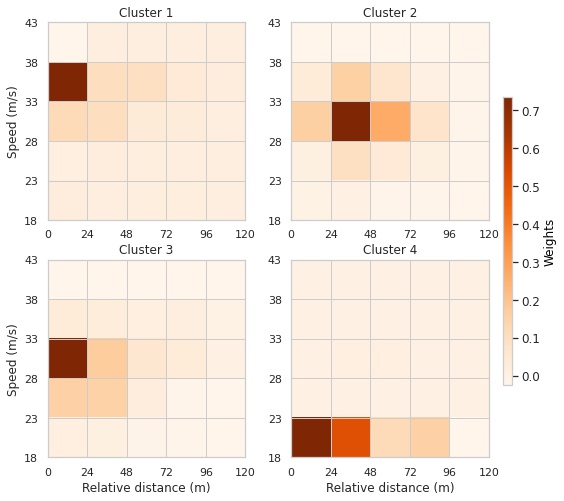}
  	\caption{Centroid of the normalized weights of the reward function for each group.}
  	\label{fig:reward}
\end{figure} 

We obtained weights of the reward function for each driver using the model-free IRL method as a 25-dimensional vector.  Then we normalized the vector into $[-1,1] $ for each driver without losing the  information of the driver's preference over different states. We divided the drivers into $K$ separate groups based on the normalized weights using K-means and the elbow method.

The goal of K-means clustering is to divide observations into $K$ clusters such that each observation belongs to the clustering with the nearest mean. The elbow method is commonly used to find an optimal number of clusters. 

Figure~\ref{fig:elbow} shows the inertia and distortion with respect to the value of $K$. Inertia is the sum of squared distances of samples to their closest cluster center, and distortion is the average of the squared distances from the cluster centers of the respective clusters using the Euclidean distance metric. Both inertia and distortion will decrease with the increase of clustering number $K$ as the sample partition becomes more refined. The decrease will be sharp before reaching the true clustering number,  and it will become flat afterward~\cite{liu2020determine}. As shown in Figure~\ref{fig:elbow}, the decrease is sharper before $K$ reaches 4 and becomes relatively flat after reaching 4. Therefore, we selected the number of representative driver groups in the car-following scenario $K$ as 4.

% \subsection{Reward function for each cluster}

Figure~\ref{fig:reward} shows the centroid of the normalized weights for each group. Drivers in Cluster 1 prefer to drive at a relatively high speed with a short distance, representing a confident driving style. The drivers in Cluster 2 and Cluster 3 prefer the state with the speed between 28  $m/s$ and 33 $m/s$, while the drivers in Cluster 2 prefer greater distance. Drivers in Cluster 4 prefer to drive at a relatively low speed and short distance, which is usually the strategy when the traffic is relatively heavy. 

\begin{figure}[htbp]
  	\centering
  	\includegraphics[width=70mm]{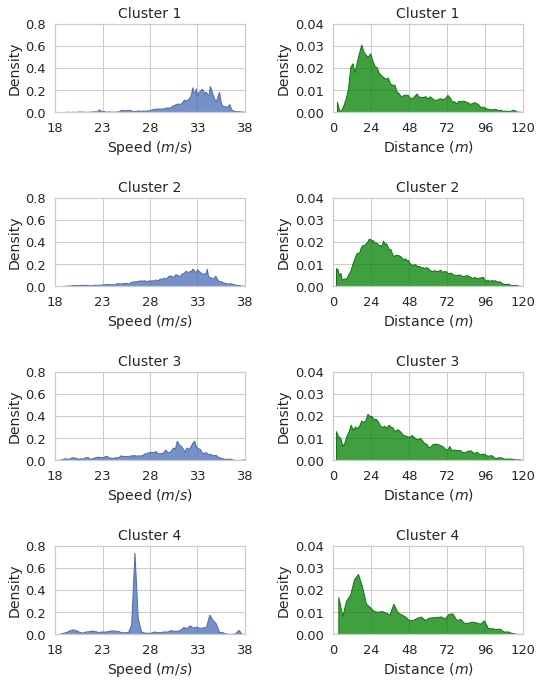}
  	\caption{ Histogram of the speed and the distance for each cluster of drivers. }
  	\label{fig:DataDistribution}
\end{figure} 

Figure~\ref{fig:DataDistribution} shows the histograms of the speed and the distance for different clusters of drivers. It is evident that the speed and the distance distributions for each cluster have different shapes, indicating different driving styles. For the speed histograms, Cluster 1 has a plateau between 32 $m/s$ and 36 $m/s$, Cluster 2 and Cluster 3   are skewed between 28 and 33 $m/s$, and Cluster 4 has a peak at 26 $m/s$. 

\begin{figure}[t]
   	\centering
   	\includegraphics[width=70mm]{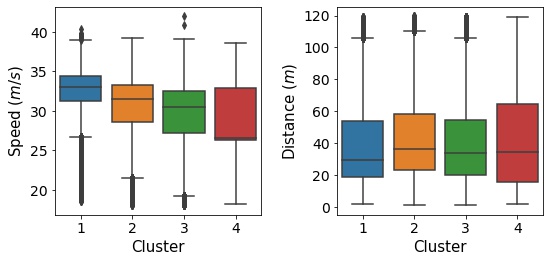}
   	\caption{Boxplot of the speed and the distance for each cluster of drivers. }
   	\label{fig:boxplot31}
\end{figure}

Figure~\ref{fig:boxplot31} shows the boxplots of the driving indicators, including the speed and the distance for the drivers in each cluster. It also demonstrates the difference between the drivers in each cluster. For the speed boxplots,  Cluster 1 has the greatest median value and  Cluster 4 has the least. Cluster 2 and cluster 3 are close. For the distance boxplots, Cluster 4 has the largest range. These characteristics are consistent with the styles represented by the reward function for each cluster.

\section{Prediction}
\label{prediction}
GMM has proven its effectiveness in modelling various driving  behaviors~\cite{Zhu2019Typical-driving-style-orientedData}. The driving data  can be modelled as a linear combination of  Gaussian distributions: 

\begin{equation}
p(x)=\sum_{i=1}^M \pi_i p(x| \mu_i, \sigma_i),    
\end{equation}

\noindent where $M$ is the number of Gaussian distribution, the $i^{th}$ component is a multivariate Gaussian distribution $G(\mu_i, \sigma_i)$ with weight  $\pi_i$. 

We considered the drivers with more than 60 car-following events as the target set, the drivers with car-following events between 11 and 60 as the source set. We further divided the car-following events in the target set into two parts: 1) Ten events for GMM modeling;  2) The rest events for validation by applying the IRL directly.

We first modeled each  driver with limited data (up to 10 events in Part 1) using a GMM $f(x)$ individually, then we modelled  each cluster of drivers in the source set into a GMM $g(x)$. For each driver in the target set, we aim to find the most similar cluster of drivers. We used KL-divergence is often used to measure the similarity.  \cite{Zhu2019Typical-driving-style-orientedData}:
\begin{equation}
    D(f||g)= \int f(x) \log \frac{f(x)}{g(x)} dx
\end{equation}
Given that the integral is not tractable, the Monte-Carlo sampling method is used to approximate the KL-divergence. The smaller the KL-divergence, the greater the similarity. In other words, we aimed to find the smallest KL-divergence.

\begin{figure}[t]
  	\centering
  	\includegraphics[width=55mm]{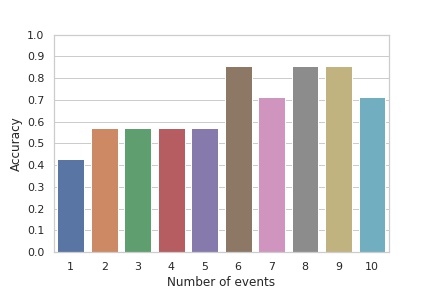}
  	\caption{Prediction accuracy.}
  	\label{accuracy}
\end{figure}

We directly learned the reward function from the events in Part 2 through the IRL, and used its closet cluster as the ground truth.  Figure~\ref{accuracy} shows the prediction accuracy. When the number of events used to build the GMM model is less than 6, the accuracy is less than 60\%. When the number increased to 6 and beyond, the accuracy also increased. The best accuracy is 85.7\% for the event numbers 6, 8, and 9. It demonstrates that our method  has the potential to predict a personalized reward function for the drivers with limited data.

\section{Online controller based on Cost-constrained POMCP}
\label{POMCP}
We designed a personalized adaptive cruise controller based on cost-constrained partially observable Monte-Carlo Planner (CC-POMCP) using the learned IRL reward.

\subsection{Preliminaries}
Formally, a POMDP is denoted as a tuple $(S, A, \mathcal{T}, R, O, \mathcal{\delta}, \gamma)$, 
where $S$ is a finite of state, $A$ is a set of actions, $\mathcal{T}$ is the transition function representing conditional transition probabilities between states, $R: S \times A \to \mathbb{R}$ is the real-valued reward function, $O$ is a set of observations, $\mathcal{\delta}$ is the observation function representing the conditional probabilities of observations given states and actions, and $\gamma \in [0,1]$ is the discount factor.   
At each time step $t$, given an action $a_t \in A$, a state $s_t \in S$ evolves to $s_{t+1} \in S$ with probability $\mathcal{T}(s_{t+1} | s_t, a_t)$. The agent receives a reward $R(s_t, a_t)$, and makes an observation $o_{t+1} \in O$ about the next state $s_{t+1}$ with probability $\delta(o_{t+1} | s_{t+1}, a_t)$. 
The goal of POMDP planning is to compute the optimal policy that chooses actions to maximize the expectation of the cumulative reward $V_R=\mathbb{E} [\sum_{t=0}^\infty \gamma^t R(s_t, a_t)]$. Constrained POMDP is a generalization of POMDP for multiple objectives. Its goal is to compute the optimal policy that maximizes $V_R$ while constraining the expected cumulative costs $V_C=\mathbb{E} [\sum_{t=0}^\infty \gamma^t C(s_t, a_t)]$, where $C(s_t, a_t)$ below a threshold $c$.
\begin{figure}%[htbp]
  	\centering
  	\includegraphics[width=55mm]{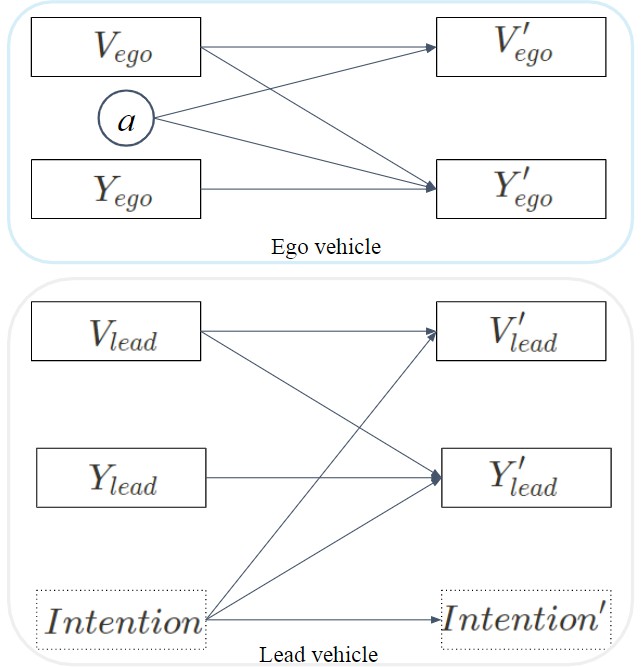}
  	\caption{Transition model.}
  	\label{pomdp}
\end{figure} 
\subsection{PACC POMDP model}
We designed a POMDP model as shown in Figure~\ref{pomdp} to simulate the driving situation that one ego vehicle following a leading vehicle on a straight highway. We used $v_{ego}$ and $y_{ego}$ to represent the speed  and  the position of the ego vehicle, respectively. We also used $v_{lead}$, $y_{lead}$, and $i_{lead}$ to represent the speed, the position of the leading vehicle. In addition, we modeled the intention of the leading vehicle's driver as a hidden state. The intention can take one of the three values, namely hesitating, normal, and aggressive.
The action $a$ in our POMDP model is the acceleration of the ego vehicle. In our experiment, the acceleration can take of the three values, namely  $-$0.6 $m/s^2$, 0, and 0.6  $m/s^2$. The state transition can be represented as below,

$$\begin{bmatrix} v'_{ego} \\ y'_{ego}\\ v'_{lead}\\ y'_{lead}  \end{bmatrix} = \begin{bmatrix} v_{ego} \\ y_{ego}\\ v_{lead}\\ y_{lead}  \end{bmatrix} +  \begin{bmatrix} a_{}t \\ v_{ego}t+0.5a_{}t^2 \\ a_{lead}t \\ v_{lead}t+0.5a_{lead}t^2  \end{bmatrix}, $$
where the time step is $t$.

We assume that the behavior of the leading vehicle is dictated by its driver's intention, as described in Table~\ref{tab:distribution}. For example, if the intention of the driver in the leading vehicle is hesitating, we assume that the probability of breaking (($a_{lead}=-0.5\;m/s^2$)) is 0.3, of maintaining ($a_{lead}=0$) is 0.4, and of accelerating ($a_{lead}=0.5\;m/s^2$) is 0.3.

\begin{table}[]\centering
\caption{Distribution of acceleration based on a different intention} 
\label{tab:distribution} 
\begin{tabular}{c|ccc}
\hline
\multirow{2}{*}{Intention} & \multicolumn{3}{c}{Acceleration} \\ \cline{2-4} 
                           & Break  & Maintain & Acceleration \\ \hline
Hesitating                 & 0.3    & 0.4      & 0.3          \\
Normal                     & 0.1    & 0.8      & 0.1          \\
Aggressive                 & 0.4    & 0.2      & 0.4          \\ \hline
\end{tabular}
\end{table}

\begin{figure}[t]
  	\centering
  	\includegraphics[width=55mm]{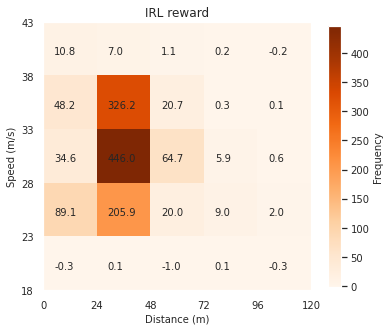}
  	\caption{The reward function for PACC.}
  	\label{rewardPACC}
\end{figure} 

We designed a reward function as shown in Figure~\ref{rewardPACC}. The reward represents drivers' driving style. In addition, we designed a cost value of 10 for the situation when the distance between the ego vehicle and leading vehicle is smaller than 2 $m$ to ensure safety.

\subsection{CC-POMCP}
\begin{figure}[t]
  	\centering
  	\includegraphics[width=60mm]{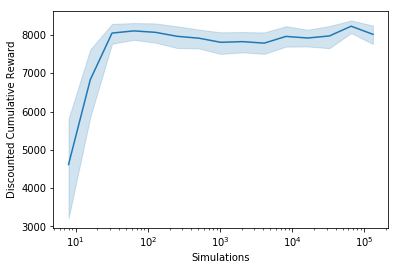}
  	\caption{The cumulative reward with respect to the number of simulations.}
  	\label{RewardVsSimuation}
\end{figure}

\begin{figure}[t]
  	\centering
  	\includegraphics[width=60mm]{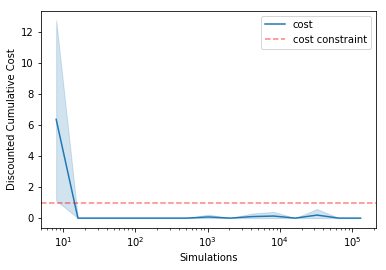}
  	\caption{The cumulative cost with respect to the number of simulations.}
  	\label{CostVsSimulation}
\end{figure}

We employed a CC-POMCP solver for our experiment ~\cite{lee2018monte}. A state is sampled from the root node's belief and then further used to sample a trajectory. More simulations tend to yield a higher cumulative reward (see Figure~\ref{RewardVsSimuation}) and lower cumulative cost (see Figure~\ref{CostVsSimulation}). However, more simulations required a longer computational time. 
% (see Figure~\ref{TimeVsSimulation})
 Since we aimed to output control commands at 1 Hz, we limited the computational time to be less than 1 second for each command output. 

\subsection{Results}
\begin{figure}[t]
  	\centering
  	\includegraphics[width=60mm]{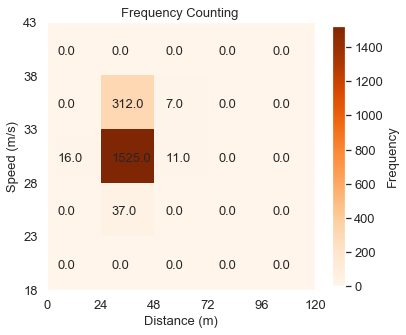}
  	\caption{The scatter plot and histogram of the speed and distance of the ego vehicle.}
  	\label{trajectoryScatter}
\end{figure}

We generate 30 car-following trajectories. In each trajectory, the ego vehicle followed the leading vehicle to the destination on a straight way. As discussed above, we set a cost constraint to ensure a safe distance. At the same time, the acceleration controlled  by the CC-POMCP solver tries to mimic the driver's driving preference. Figure~\ref{trajectoryScatter} shows the scatter plot of the speed and distance of all the 30 trajectories, and their histogram, respectively. It shows that the trajectories are mostly scattered in the states with relatively high rewards as shown in Figure~\ref{rewardPACC}. 

\begin{figure}[t]
  	\centering
  	\includegraphics[width=60mm]{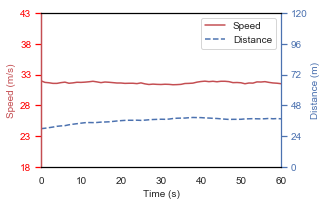}
  	\caption{The average trajectory of the ego vehicle from 30 runs of the experiment.}
  	\label{trajectory_PACC}
\end{figure}

As shown in Figure~\ref{trajectory_PACC}, the average trajectory of the 30 trajectories is relatively stable.  The speed maintains between 28 $m/s$ and 33 $m/s$, and the relative distance maintains between 24 $m$ and $48$, which is the most preferred state in Figure~\ref{rewardPACC}. 

\section{Discussion and Conclusion}

\label{conclustion}
This work proposed a method to learn personalized driving styles for drivers with limited data. We first extracted car-following events from a naturalistic driving dataset, and divided them into a source set and a target set. We adopted a model-free IRL to learn the driving styles through reward function in the car-following scenarios for the events in the source set. We further clustered these driving styles into 4 representative groups. However, the IRL  requires enough data and it is not always available for newly-involved drivers. We used GMM for the driver style in the target set using up to 10 events. Then  we used KL-divergence to find the most similar cluster in the source set. 
To validate our method, we used more than 50 events in the target set to learn the reward function as the ground truth to verify whether the prediction-based KL-divergence is valid or not. We achieved a prediction accuracy of 85.7\% in this study. Furthermore, we employed a CC-POMCP solver for the control of the ego vehicle in a car-following experiment. By incorporating the reward function learned from drivers, we are able to mimic drivers' driving styles and provide a personalized driving experience.

There are a few directions for future work. First, we would like to expand intervals of the states and incorporate more features. We believe it will provide a more comprehensive analysis of drivers' preferences. Another important future work is to adapt the personalized model to a more complex driving situation.

\section*{Acknowledgement}
This work was supported in part by National Science Foundation grants CCF-1942836, CNS-1755784, and Toyota Motor North America ``Digital Twins'' project.
Any opinions, findings, and conclusions or recommendations expressed in this material are those of the author(s) and do not necessarily reflect the views of the grant sponsors.

\printbibliography 

@article{Wang2018AModel,
    title = {{A learning-based approach for lane departure warning systems with a personalized driver model}},
    year = {2018},
    journal = {IEEE Transactions on Vehicular Technology},
    author = {Wang, Wenshuo and Zhao, Ding and Han, Wei and Xi, Junqiang},
    number = {10},
    pages = {9145--9157},
    volume = {67}
}

@article{Wang2019AModels,
    title = {{A Learning-Based Personalized Driver Model Using Bounded Generalized Gaussian Mixture Models}},
    year = {2019},
    journal = {IEEE Transactions on Vehicular Technology},
    author = {Wang, Wenshuo and Xi, Junqiang and Hedrick, J. Karl},
    number = {12},
    pages = {11679--11690},
    volume = {68},
    publisher = {IEEE}
}

@article{Yi2019ARecognition,
title={A machine learning based personalized system for driving state recognition},
  author={Yi, Dewei and Su, Jinya and Liu, Cunjia and Quddus, Mohammed and Chen, Wen-Hua},
  journal={Transportation Research Part C: Emerging Technologies},
  volume={105},
  pages={241--261},
  year={2019},
  publisher={Elsevier}
}

@inproceedings{abbeel2004apprenticeship,
  title={Apprenticeship learning via inverse reinforcement learning},
  author={Abbeel, Pieter and Ng, Andrew Y},
  booktitle={Proceedings of the twenty-first international conference on Machine learning},
  pages={1},
  year={2004}
}

@article{Kuderer2015LearningDemonstration,
    title = {{Learning driving styles for autonomous vehicles from demonstration}},
    year = {2015},
    journal = {Proceedings - IEEE International Conference on Robotics and Automation},
    author = {Kuderer, Markus and Gulati, Shilpa and Burgard, Wolfram},
    number = {June},
    pages = {2641--2646},
    volume = {2015-June},
    publisher = {IEEE},

}

@inproceedings{Jain2019Model-freeEstimation,
  title={Model-free IRL using maximum likelihood estimation},
  author={Jain, Vinamra and Doshi, Prashant and Banerjee, Bikramjit},
  booktitle={Proceedings of the AAAI Conference on Artificial Intelligence},
  volume={33},
  number={01},
  pages={3951--3958},
  year={2019}
}

@article{Bolduc2019MultimodelControl,
    title = {{Multimodel approach to personalized autonomous adaptive cruise control}},
    year = {2019},
    journal = {IEEE Transactions on Intelligent Vehicles},
    author = {Bolduc, Andrew Phillip and Guo, Longxiang and Jia, Yunyi},
    number = {2},
    pages = {321--330},
    volume = {4},
    publisher = {IEEE}
}

@article{Gao2020PersonalizedControl,
  title={Personalized adaptive cruise control based on online driving style recognition technology and model predictive control},
  author={Gao, Bingzhao and Cai, Kunyang and Qu, Ting and Hu, Yunfeng and Chen, Hong},
  journal={IEEE transactions on vehicular technology},
  volume={69},
  number={11},
  pages={12482--12496},
  year={2020},
  publisher={IEEE}
}

@article{Zhu2018PersonalizedIdentification,
    title = {{Personalized Lane-Change Assistance System with Driver Behavior Identification}},
    year = {2018},
    journal = {IEEE Transactions on Vehicular Technology},
    author = {Zhu, Bing and Yan, Shude and Zhao, Jian and Deng, Weiwen},
    number = {11},
    pages = {10293--10306},
    volume = {67},
    publisher = {IEEE}
}

@article{Zhu2019Typical-driving-style-orientedData,
  title={Typical-driving-style-oriented Personalized Adaptive Cruise Control design based on human driving data},
  author={Zhu, Bing and Jiang, Yuande and Zhao, Jian and He, Rui and Bian, Ning and Deng, Weiwen},
  journal={Transportation research part C: emerging technologies},
  volume={100},
  pages={274--288},
  year={2019},
  publisher={Elsevier}
}

@phdthesis{vroman2014maximum,
  title={Maximum likelihood inverse reinforcement learning},
  author={Vroman, Monica C},
  year={2014},
  school={Rutgers University-Graduate School-New Brunswick}
}

@inproceedings{ramachandran2007bayesian,
  title={Bayesian Inverse Reinforcement Learning.},
  author={Ramachandran, Deepak and Amir, Eyal},
  booktitle={IJCAI},
  volume={7},
  pages={2586--2591},
  year={2007}
}

@article{liu2020determine,
  title={Determine the number of unknown targets in Open World based on Elbow method},
  author={Liu, Fan and Deng, Yong},
  journal={IEEE Transactions on Fuzzy Systems},
  year={2020},
  publisher={IEEE}
}

@article{hasenjager2019survey,
  title={A Survey of Personalization for Advanced Driver Assistance Systems},
  author={Hasenj{\"a}ger, Martina and Heckmann, Martin and Wersing, Heiko},
  journal={IEEE Transactions on Intelligent Vehicles},
  volume={5},
  number={2},
  pages={335--344},
  year={2019},
  publisher={IEEE}
}

@article{lu2019transfer,
  title={Transfer Learning for Driver Model Adaptation in Lane-Changing Scenarios Using Manifold Alignment},
  author={Lu, Chao and Hu, Fengqing and Cao, Dongpu and Gong, Jianwei and Xing, Yang and Li, Zirui},
  journal={IEEE Transactions on Intelligent Transportation Systems},
  year={2019},
  publisher={IEEE}
}

@inproceedings{li2019transferable,
  title={Transferable Driver Behavior Learning via Distribution Adaption in the Lane Change Scenario},
  author={Li, Zirui and Gong, Cheng and Lu, Chao and Gong, Jianwei and Lu, Junyan and Xu, Youzhi and Hu, Fengqing},
  booktitle={2019 IEEE Intelligent Vehicles Symposium (IV)},
  pages={193--200},
  year={2019},
  organization={IEEE}
}

@article{bagloee2016autonomous,
  title={Autonomous vehicles: challenges, opportunities, and future implications for transportation policies},
  author={Bagloee, Saeed Asadi and Tavana, Madjid and Asadi, Mohsen and Oliver, Tracey},
  journal={Journal of modern transportation},
  volume={24},
  number={4},
  pages={284--303},
  year={2016},
  publisher={Springer}
}

@article{gao2020personalized,
  title={Personalized Adaptive Cruise Control Based on Online Driving Style Recognition Technology and Model Predictive Control},
  author={Gao, Bingzhao and Cai, Kunyang and Qu, Ting and Hu, Yunfeng and Chen, Hong},
  journal={IEEE Transactions on Vehicular Technology},
  year={2020},
  publisher={IEEE}
}

@inproceedings{chen2017learning,
  title={A learning model for personalized adaptive cruise control},
  author={Chen, Xin and Zhai, Yong and Lu, Chao and Gong, Jianwei and Wang, Gang},
  booktitle={2017 IEEE Intelligent Vehicles Symposium (IV)},
  pages={379--384},
  year={2017},
  organization={IEEE}
}

@inproceedings{shaout2011advanced,
  title={Advanced driver assistance systems-past, present and future},
  author={Shaout, Adnan and Colella, Dominic and Awad, SS},
  booktitle={2011 Seventh International Computer Engineering Conference (ICENCO'2011)},
  pages={72--82},
  year={2011},
  organization={IEEE}
}

@inproceedings{ng2000algorithms,
  title={Algorithms for inverse reinforcement learning.},
  author={Ng, Andrew Y and Russell, Stuart J and others},
  booktitle={ICML},
  volume={1},
  pages={2},
  year={2000}
}

@inproceedings{naumann2020analyzing,
  title={Analyzing the Suitability of Cost Functions for Explaining and Imitating Human Driving Behavior based on Inverse Reinforcement Learning},
  author={Naumann, Maximilian and Sun, Liting and Zhan, Wei and Tomizuka, Masayoshi},
  booktitle={2020 IEEE International Conference on Robotics and Automation (ICRA)},
  pages={5481--5487},
  year={2020},
  organization={IEEE}
}

@article{taubman2004multidimensional,
  title={The multidimensional driving style inventory—scale construct and validation},
  author={Taubman-Ben-Ari, Orit and Mikulincer, Mario and Gillath, Omri},
  journal={Accident Analysis \& Prevention},
  volume={36},
  number={3},
  pages={323--332},
  year={2004},
  publisher={Elsevier}
}

@inproceedings{silver2013learning,
  title={Learning autonomous driving styles and maneuvers from expert demonstration},
  author={Silver, David and Bagnell, J Andrew and Stentz, Anthony},
  booktitle={Experimental Robotics},
  pages={371--386},
  year={2013},
  organization={Springer}
}

@inproceedings{levine2012continuous,
author = {Levine, Sergey and Koltun, Vladlen},
title = {Continuous Inverse Optimal Control with Locally Optimal Examples},
year = {2012},


booktitle = {Proceedings of the 29th International Coference on International Conference on Machine Learning},
pages = {475–482},
numpages = {8},
series = {ICML'12}
}

@inproceedings{sun2018probabilistic,
  title={Probabilistic prediction of interactive driving behavior via hierarchical inverse reinforcement learning},
  author={Sun, Liting and Zhan, Wei and Tomizuka, Masayoshi},
  booktitle={2018 21st International Conference on Intelligent Transportation Systems (ITSC)},
  pages={2111--2117},
  year={2018},
  organization={IEEE}
}

@INPROCEEDINGS{rosbach2019driving,
  author={S. {Rosbach} and V. {James} and S. {Großjohann} and S. {Homoceanu} and S. {Roth}},
  booktitle={2019 IEEE/RSJ International Conference on Intelligent Robots and Systems (IROS)}, 
  title={Driving with Style: Inverse Reinforcement Learning in General-Purpose Planning for Automated Driving}, 
  year={2019},
  volume={},
  number={},
  pages={2658-2665},
}

@article{arora2021survey,
  title={A survey of inverse reinforcement learning: Challenges, methods and progress},
  author={Arora, Saurabh and Doshi, Prashant},
  journal={Artificial Intelligence},
  pages={103500},
  year={2021},
  publisher={Elsevier}
}

@article{uchibe2018model,
  title={Model-free deep inverse reinforcement learning by logistic regression},
  author={Uchibe, Eiji},
  journal={Neural Processing Letters},
  volume={47},
  number={3},
  pages={891--905},
  year={2018},
  publisher={Springer}
}

@inproceedings{ziebart2008maximum,
  title={Maximum entropy inverse reinforcement learning.},
  author={Ziebart, Brian D and Maas, Andrew L and Bagnell, J Andrew and Dey, Anind K},
  booktitle={AAAI},
  volume={8},
  pages={1433--1438},
  year={2008},

}

@inproceedings{shiarlis2016inverse,
  title={Inverse reinforcement learning from failure},
  author={Shiarlis, Kyriacos and Messias, Joao and Whiteson, SA},
  year={2016},
  booktitle={International Foundation for Autonomous Agents and Multiagent Systems},
   pages={1060--1068},
}

@ARTICLE{wang2020driver,  author={Ziran {Wang} and X. {Liao} and C. {Wang} and D. {Oswald} and G. {Wu} and K. {Boriboonsomsin} and M. {Barth} and K. {Han} and B. {Kim} and P. {Tiwari}},  journal={IEEE Transactions on Intelligent Vehicles},   title={Driver Behavior Modeling using Game Engine and Real Vehicle: A Learning-Based Approach},   year={2020},  volume={5},  number={4},  pages={738-749},}

@ARTICLE{wang2020asurvey,  author={Ziran {Wang} and Y. {Bian} and S. E. {Shladover} and G. {Wu} and S. E. {Li} and M. J. {Barth}},  journal={IEEE Intelligent Transportation Systems Magazine},   title={A Survey on Cooperative Longitudinal Motion Control of Multiple Connected and Automated Vehicles},   year={2020},  volume={12},  number={1},  pages={4-24},}

@inproceedings{lee2018monte,
  title={Monte-Carlo Tree Search for Constrained POMDPs.},
  author={Lee, Jongmin and Kim, Geon-Hyeong and Poupart, Pascal and Kim, Kee-Eung},
  booktitle={NeurIPS},
  pages={7934--7943},
  year={2018}
}

@inproceedings{wang2021personalized,
  title={Personalized Adaptive Cruise Control via Gaussian Process Regression},
  author={Wang, Yanbing and Wang, Ziran and Han, Kyungtae and Tiwari, Prashant and Work, Daniel B},
  booktitle={2021 IEEE International Intelligent Transportation Systems Conference (ITSC)},
  pages={1496--1502},
  year={2021},
}

@article{gao2018car,
  title={Car-following method based on inverse reinforcement learning for autonomous vehicle decision-making},
  author={Gao, Hongbo and Shi, Guanya and Xie, Guotao and Cheng, Bo},
  journal={International Journal of Advanced Robotic Systems},
  volume={15},
  number={6},
  pages={1729881418817162},
  year={2018},
  publisher={SAGE Publications Sage UK: London, England}
}

@inproceedings{zhao2022personalized,
  title={Personalized Car Following for Autonomous Driving with Inverse Reinforcement Learning},
  author={Zhao, Zhouqiao and Wang, Ziran and Han, Kyungtae and Tiwari, Prashant and Wu, Guoyuan and Barth, Matthew},
  booktitle={Proceedings 2022 IEEE International Conference on Robotics and Automation},
  year={2022},
}

@inproceedings{liao2022online,
  title={Online Prediction of Lane Change with a Hierarchical Learning-Based Approach},
  author={Liao, Xishun and Wang, Ziran and Zhao, Xuanpeng and Zhao, Zhouqiao and Han, Kyungtae and Tiwari, Prashant and Barth, Matthew and Wu, Guoyuan},
  booktitle={Proceedings 2022 IEEE International Conference on Robotics and Automation},
  year={2022},
  organization={IEEE}
}
\end{document}